\documentclass[11pt]{article}

\usepackage{amsmath}
\usepackage{mathtools}
\usepackage{amssymb}
\usepackage{amsfonts}
\usepackage{amsthm}
\usepackage{mathrsfs}
\usepackage[all]{xy}
\usepackage[colorlinks=true, citecolor=blue]{hyperref}
\usepackage{tensor}

\usepackage{authblk}

\newcommand{\dd}{\mathrm{d}}

\def\Horava{Ho\v{r}ava }

\begin{document}

\title{Comment on ``Scalar Einstein-Aether theory"}
\author[1,2]{Ted Jacobson\thanks{jacobson@umd.edu}}
\author[1,2]{Antony J. Speranza\thanks{asperanz@umd.edu}}

\affil[1]{\small \it Maryland Center for Fundamental Physics, University of Maryland, \mbox{College Park,} 
MD 20742, USA}
\affil[2]{\small \it Perimeter Institute for Theoretical Physics, 31 Caroline Street North, Waterloo, ON N2L 2Y5, 
Canada}
%
\date{June 12, 2014}
\maketitle


A recent paper \cite{Haghani2014a} 
studies a modification of Einstein-aether theory  \cite{Jacobson2001} in which the aether vector
is restricted, at the level of the action, to be the gradient of a scalar, $u_a = \nabla_a S$,
constrained to be a timelike unit vector. In this 
comment  we note that 
this scalar version of Einstein-aether theory,
referred to here as $S$-theory for brevity,
is equivalent to the projectable version of the IR limit of Ho\v{r}ava gravity \cite{Horava2009}
when the potential $V(S)$ is constant. 
The covariant formulation of projectable \Horava gravity provided by $S$-theory was first described by 
\cite{Blas2009}. In this comment we briefly explain the relation to nonprojectable \Horava gravity and  to 
Einstein-aether theory.

It has previously been shown \cite{Blas2009,Jacobson2010a} that the 
nonprojectable version of IR-Ho\v{r}ava gravity 
is obtained from the Einstein-aether action with the aether required
to be hypersurface orthogonal. That is, in the action the aether
is of the form $u_a = N\nabla_a T$,
where $T$ is a scalar field and  $N$ is chosen so that $u_a$ has unit norm.
We refer to this as $T$-theory.  The unit norm constraint may be implemented
either by setting $N=(g^{ab}\nabla_a T\nabla_b T)^{-1/2}$ in the action,
or by a constraint term $\lambda(u^au_a - 1)$ with Lagrange multiplier 
$\lambda$.
The scalar $T$ serves to define the preferred 
foliation that is central to Ho\v{r}ava gravity, and $N$ corresponds to the lapse
function. Since the $T$ equation of motion follows from the other equations of motion,
$T$ may be set equal to the coordinate time in the action, resulting in 
the action for nonprojectable IR-\Horava  gravity \cite{Jacobson2010a}.

 $S$-theory is equivalent to $T$-theory with the further
restriction that $N\, \dd T = \dd S$ for some scalar field $S$. This restriction implies 
that $N=N(T)$, i.e.\ $N$ depends on spacetime only via $T$, which is precisely the 
projectability condition for the lapse function. The unit constraint then cannot be 
solved by a choice of $N$, since for a generic $T$ the required $N$ would not be 
just a function of $T$, but rather must 
be imposed by the constraint term $\lambda(\nabla^aS \nabla_a S - 1)$.
In this formulation of projectable \Horava gravity, the $T$ 
reparameterization symmetry
under which $\dd S=N\dd T$ is invariant reduces to the shift symmetry $S\rightarrow S+\rm{const}$.

The general action considered in Ref.\ \cite{Haghani2014a} includes a potential term $V(S)$.  
This is incompatible with the $S$ shift symmetry
unless $V$ is constant, in which case it is just a cosmological constant.  
When a non-constant potential of this sort is included 
the theory is therefore no longer dynamically equivalent to projectable IR Ho\v{r}ava gravity. 

\subsection*{Acknowledgements}
This research was supported in part by the NSF under grant No.\ PHY-0903572 and 
by Perimeter Institute for Theoretical Physics.  Research at Perimeter Institute is supported by the 
Government of Canada through Industry Canada and by the Province of Ontario through the Ministry of 
Research \& Innovation. 

\bibliographystyle{abbrvunsrturl}
\bibliography{comment_on_scalar_aetheory.bib}

\end{document}